%% file: bare_conf.tex
\documentclass[conference,letterpaper]{IEEEtran}

\usepackage{cite}
\usepackage{amsmath,amssymb,amsfonts}
\usepackage{algorithmic}
\usepackage{acronym}
\usepackage{textcomp}
\usepackage{subcaption}
\usepackage{float}
\usepackage{array}
\usepackage{soul, color, acronym, multirow, balance}
\usepackage[pdftex]{graphicx}
\usepackage{epsfig, epstopdf}
\usepackage{booktabs}
\usepackage[hyphens]{url}
\usepackage[justification=centering]{caption}

\usepackage[colorinlistoftodos]{todonotes}

\input{sections/acr.tex}

\begin{document}
\title{Exploring Machine Learning for Classification of QUIC Flows over Satellite}
\author{
		\IEEEauthorblockN{
		 Raffaello Secchi\IEEEauthorrefmark{1},
		Pietro Cassar\`a\IEEEauthorrefmark{2}, and Alberto Gotta\IEEEauthorrefmark{2}}
		\IEEEauthorrefmark{2}School of Engineering, University of Aberdeen, UK\\
		\IEEEauthorrefmark{2}National Research Council (CNR), Information Science and Technologies Institute (ISTI), Pisa, Italy\\ 

	}
\maketitle

\begin{abstract} Automatic traffic classification is increasingly important in networking due to the current trend of encrypting transport information (e.g., behind HTTP encrypted tunnels) which prevent intermediate nodes to access end-to-end transport headers. This paper proposes an architecture for supporting Quality of Service (QoS) in hybrid terrestrial and SATCOM networks based on automated traffic classification. Traffic profiles are constructed by machine-learning (ML) algorithms using the series of packet sizes and arrival times of QUIC connections. Thus, the proposed QoS method does not require explicit setup of a path (i.e. it provides soft QoS), but  employs agents within the network to verify that flows conform to a given traffic profile. Results over a range of ML models encourage integrating ML technology in SATCOM equipment. The availability of higher computation power at low-cost creates the fertile ground for implementation of these techniques.
\end{abstract}

\IEEEpeerreviewmaketitle

\input{sections/Introduction.tex}

\input{sections/RelatedWorks.tex}
\input{sections/SystemArchitecture.tex}

\input{sections/ProblemDefinition.tex}

\input{sections/PerformanceEvaluation.tex}
\input{sections/FeaturesAnalysis.tex}

\section{Conclusion}
\label{sec:conclusions}

We proposed a QoS architecture where end-hosts connected to the satellite network forward traffic over available paths and internal nodes verify that traffic is conforming to the characteristic of the path. Our goal is not to determine the exact application originating the traffic (i.e. an exact fingerprinting), but rather to categorise the traffic into classes (e.g. real-time, quasi real-time, and delay tolerant) that can receive similar QoS treatment. For example, if the flow is characterised by small packets at regular intervals, it could be identified as a real-time flow (such as audio or gaming) and should receive a specific low-latency treatment.

The results show that several ML learning algorithms can generalise performance to the satellite network even if they were trained in non-satellite conditions.  The results can be further strengthen by selecting a more descriptive set of features using information metrics. All this analysis is an encouraging to integrate ML classification within SATCOM technology.





\section*{Acknowledgment}
This work is partially funded by the European Space Agency, ESA-ESTEC, Noordwijk, The Netherlands, under contract n. 4000130962/20/NL/NL/FE (“SatNEx V— Satellite Network of Experts V”). The view expressed herein can in no way be taken to reflect the official opinion of the European Space Agency.

\IEEEtriggeratref{0}
\balance
\bibliographystyle{IEEEtran}
\bibliography{references}
\end{document}

%% file: sections/acr.tex
\acrodef{PER}{Packet Erasure Rate}
\acrodef{DPI}{Deep Packet Inspection}
\acrodef{ML}{Machine Learning}
\acrodef{DL}{Deep Learning}
\acrodef{FL}{Federated Learning}
\acrodef{PLR}{Packet Loss Rate}
\acrodef{PRR}{Packet Received Ratio}
\acrodef{BER}{Bit Error Rate}
\acrodef{SG}{Smart Gateway}
\acrodef{SS}{Smart Switch}
\acrodef{RTT}{Round-Trip Time}
\acrodef{TCP}{Transmission Control Protocol}
\acrodef{UDP}{User Datagram Protocol}
\acrodef{AIMD}{Additive Increase Multiplicative Decrease}
\acrodef{PEP}{Performance Enhancing Proxy}
\acrodef{cwnd}{congestion window}
\acrodef{STP}{Satellite Transport Protocol}
\acrodef{BDP}{Bandwidth-Delay Product}
\acrodef{QoS}{Quality of Service}
\acrodef{LoS}{Line of Sight}
\acrodef{NLoS}{Non Line of Sight}
\acrodef{BLoS}{Beyond Line of Sight}
\acrodef{QoE}{Quality of Experience}
\acrodef{ACM}{Adaptive Coding and Modulation}
\acrodef{DAMA}{Demand Assignment Multiple Access}
\acrodef{VANET}{Vehicular Ad-Hoc Network}
\acrodef{RA}{Random Access}
\acrodef{CRA}{Contention Resolution ALOHA}
\acrodef{SA}{Slotted ALOHA}
\acrodef{DSA}{Diversity Slotted ALOHA}
\acrodef{OBU}{On-Board Unit}
\acrodef{CRDSA}{Contention Resolution Diversity Slotted ALOHA}
\acrodef{SIC}{Successive Interference Cancellation}
\acrodef{ARQ}{Automatic Repeat reQuest}
\acrodef{SC-ARQ}{Selective-Coded ARQ}
\acrodef{SR-ARQ}{Selective-Repeat ARQ}
\acrodef{IRSA}{Irregular Repetition Slotted ALOHA}
\acrodef{CGC}{Complementary Ground Component}
\acrodef{GCS}{Ground Control Station}
\acrodef{RSU}{Road Side Unit}
\acrodef{ACK}{Acknowledgment}
\acrodef{NACK}{Negative Acknowledgment}
\acrodef{DVB-SH}{Digital Video Broadcasting - Satellite Services to Handhelds}
\acrodef{DVB-H}{Digital Video Broadcasting - Handheld}
\acrodef{DVB-RCS2}{Digital Video Broadcasting - Return Channel via Satellite, II generation}
\acrodef{SACK}{Selective Acknowledgment}
\acrodef{SNACK}{Selective Negative Acknowledgment}\acrodef{SNACK}{Selective Negative Acknowledgment}
\acrodef{SNIR}{Signal to Noise plus Interference Ratio}
\acrodef{SCPS-TP}{Space Communications Protocol Specifications - Transport Protocol}
\acrodef{CCSDS}{Consultative Committee for Space Data Systems}
\acrodef{ESA}{European Space Agency}
\acrodef{NASA}{National Aeronautics and Space Administration}
\acrodef{BSM}{Broadband Satellite Multimedia}
\acrodef{RLNC}{Random Linear Network Coding}
\acrodef{NC}{Network Coding}
\acrodef{FIFO}{First In, First Out}
\acrodef{FCFS}{First Come, First Served}
\acrodef{BLER}{Block Error Rate}
\acrodef{GEO}{geosynchronous}
\acrodef{PMF}{Probability Mass Function}
\acrodef{LEO}{Low Earth Orbit}
\acrodef{FTP}{File Transfer Protocol}
\acrodef{CRC}{Cyclic Redundancy Check}
\acrodef{MAC}{Media Access Control}
\acrodef{PHY}{Physical layer}
\acrodef{HTTP}{Hypertext Transfer Protocol}
\acrodef{ISP}{Internet Service Provider}
\acrodef{MSS}{Maximum Segment Size}
\acrodef{BIC}{Binary Increase Congestion control}
\acrodef{AQM}{Active Queue Management}
\acrodef{XCP}{eXplicit Control Protocol}
\acrodef{ECN}{Explicit Congestion Notification}
\acrodef{CA}{Congestion Avoidance}
\acrodef{RED}{Random Early Detection}
\acrodef{TD}{Triple-Duplicate}
\acrodef{TO}{TimeOut}
\acrodef{IP}{Internet Protocol}
\acrodef{WMN}{Wireless Mesh Network}
\acrodef{ssthresh}{Slow-Start threshold}
\acrodef{MPE-IFEC}{Multi Protocol Encapsulation - Inter-burst Forward Error Correction}
\acrodef{FEC}{Forward Error Correction}
\acrodef{CRC}{Cyclic Redundancy Check}
\acrodef{P2P}{Peer-to-Peer}
\acrodef{FMT}{Fade Mitigation Technique}
\acrodef{SGD}{Smart Gateway Diversity}
\acrodef{NCC}{Network Control Centre}
\acrodef{ModCod}{Modulation and Coding}
\acrodef{FIFO}{First-In-First-Out}
\acrodef{WRR}{Weighted Round Robin}
\acrodef{WFQ}{Weighted Fair Queuing}
\acrodef{NS}{Network Simulator}
\acrodef{GSE}{Generic Stream Encapsulation}
\acrodef{PDF}{Probability Density Function}
\acrodef{CDF}{Cumulative Density Function}
\acrodef{CoV}{Coefficient of Variation}
\acrodef{MSC}{Message Sequence Chart}
\acrodef{ESA}{European Space Agency}
\acrodef{LIU}{Lebanese International University}
\acrodef{TUM}{Technical University of Munich}
\acrodef{MSCE-CS}{Master of Science in Communications Engineering - Communications Systems}
\acrodef{DLR}{German Aerospace Center}
\acrodef{NOS}{Network Operating System}
\acrodef{NFs}{Network Functionalities}
\acrodef{NC-SGD}{Network Coding for SGD}
\acrodef{ATSP}{Advanced Transport Satellite Protocol}
\acrodef{STP}{Satellite Transport Protocol}
\acrodef{WMN}{Wireless Mesh Network}
\acrodef{SNR}{Signal-to-Noise Ratio}
\acrodef{SINR}{Signal-to-Interference-plus-Noise Ratio}
\acrodef{LMS}{Land Mobile Satellite}
\acrodef{LTE}{Long-Term Evolution}
\acrodef{M2M}{machine-to-machine}
\acrodef{IoT}{Internet of Things}
\acrodef{IoRT}{Internet of Remote Things}
\acrodef{RA}{Random Access}
\acrodef{UAV}{Unmanned Aerial Vehicle}
\acrodef{UTV}{Unmanned Terrestrial Vehicle}
\acrodef{UAS}{Unmanned Aerial System}
\acrodef{FANET}{Flying Ad-Hoc Network}
\acrodef{MANET}{Mobile Ad-Hoc Network}
\acrodef{VANET}{Vehicle Ad-Hoc Network}
\acrodef{C2}{Command and Control}
\acrodef{DTN}{Delay Tolerant Network}
\acrodef{COTS}{Commercial Off-the-Shelf}
\acrodef{IETF}{Internet Engineering Task Force}
\acrodef{CoAP}{Constrained Application Protocol}
\acrodef{MQTT}{Message Queuing Telemetry Transport}
\acrodef{URI}{Uniform Resource Identifier}
\acrodef{PUB/SUB}{Publish / Subscribe}
\acrodef{RCST}{Return Channel Satellite Terminal}
\acrodef{TDMA}{Time Division Multiple Access}
\acrodef{FDMA}{Frequency Division Multiple Access}
\acrodef{TCDMA}{Turbo Code Division Multiple Access}
\acrodef{PDMA}{Power Division Multiple Access}
\acrodef{WSN}{Wireless Sensor Network}
\acrodef{REST}{Representational State Transfer}
\acrodef{EDGE}{Enhanced Data rates for GSM Evolution}
\acrodef{UMTS}{Universal Mobile Telecommunications System}
\acrodef{LTE}{Long-Term Evolution}
\acrodef{E2E}{End-to-End}
\acrodef{3WHS}{Three-way Handshake}
\acrodef{SCADA}{Supervisory Control And Data Acquisition}
\acrodef{SOA}{Service-Oriented Architecture}
\acrodef{WebRTC}{Web Real-Time Communications}
\acrodef{fps}{frames per second}
\acrodef{SSIM}{Structural SIMilarity}
\acrodef{PSNR}{Peak signal-to-noise ratio}
\acrodef{RPi}{Raspberry Pi}
\acrodef{NFV}{Network Function Virtualization}
\acrodef{OPEX}{Operating Expenditures}
\acrodef{CAPEX}{Capital Expenditures}
\acrodef{MIMO}{multiple-input and multiple-output}
\acrodef{SDN}{Software Defined Network}
\acrodef{VNF}{Virtual Network Function}
\acrodef{MEC}{Mobile Edge Computing}
\acrodef{MTC}{Machine-type Communications}
\acrodef{mMTC}{Massive Machine-type Communications}
\acrodef{HTC}{Human-type Communications}
\acrodef{D2D}{device to device}
\acrodef{IIoT}{industrial IoT}
\acrodef{LPWAN}{Low-Power Wide-Area Network}
\acrodef{CPS}{Cyber-Physical System}
\acrodef{ICT}{Information and Communication Technologies}
\acrodef{SDR}{Software Defined Radio}
\acrodef{GPS}{Global Positioning System}
\acrodef{H2M}{human-to-machine}
\acrodef{MC}{Machine}
\acrodef{LA}{Local Area}
\acrodef{WA}{Wide Area}
\acrodef{HAP}{High Altitude Platform}
\acrodef{LAP}{Low Altitude Platform}
\acrodef{RAN}{Radio Access Network}
\acrodef{EIRP}{Equivalent Isotropically Radiated Power}
\acrodef{LT}{Logistics and Transportations}
\acrodef{eNB}{Evolved Node B}
\acrodef{CDN}{Content Distribution Network}
\acrodef{ITU}{International Telecommunication Union}
\acrodef{SIN}{Space Information Network}
\acrodef{DTA}{Delay Tolerant Application}
\acrodef{DSA}{Delay Sensitive Application}
\acrodef{DTN}{Delay Tolerant Networking}
\acrodef{ISL}{Inter-Satellite Link}
\acrodef{TFRC}{TCP Friendly Rate Control}
\acrodef{RACH}{Random Access CHannel}
\acrodef{LPWAN}{Low-Power Wide Area Network}
\acrodef{VRT}{Variable Rate Technology}
\acrodef{HAP}{High Altitude Platform}
\acrodef{LALE}{Low Altitude Long Endurance}
\acrodef{LASE}{Low Altitude Short Endurance}
\acrodef{MALE}{Medium Altitude Long Endurance}
\acrodef{HALE}{High Altitude Long Endurance}
\acrodef{OTA}{Over The Air}
\acrodef{TLS}{Transport Layer Security}
\acrodef{SSL}{Secure Sockets Layer}

\acrodefplural{RSU}[RSUs]{Road Side Units}
\acrodefplural{VANET}[VANETs]{Vehicular Ad-Hoc Networks}
\acrodefplural{OBU}[OBUs]{On-Board Units}
\acrodefplural{UAV}[UAVs]{Unmanned Aerial Vehicles}
\acrodefplural{FANET}[FANETs]{Flying Ad-Hoc Networks}
\acrodefplural{MANET}[MANETs]{Mobile Ad-Hoc Networks}
\acrodefplural{VANET}[VANETs]{Vehicle Ad-Hoc Networks}
\acrodefplural{UAS}[UASs]{Unmanned Aerial Systems}
\acrodefplural{HAP}[HAPs]{High Altitude Platforms}
\acrodefplural{LALE}[LALEs]{Low Altitude Long Endurance}
\acrodefplural{LASE}[LASEs]{Low Altitude Short Endurance}
\acrodefplural{MALE}[MALEs]{Medium Altitude Long Endurance}
\acrodefplural{HALE}[HALEs]{High Altitude Long Endurance}
\acrodefplural{RCST}[RCSTs]{Return Channel Satellite Terminals}
\acrodefplural{VNF}[VNFs]{Virtual Network Functions}

%% file: sections/Introduction.tex
\section{Introduction}
\label{sec:introduction}


The recently standardised QUIC protocol (Quick UDP Internet Connections) is set to replace the traditional HTTP-over-TCP web architecture~\cite{rfc9000}.
QUIC introduces a single connection to multiplex data streams carrying different parts of a web page (a multi-streaming protocol). This provides great flexibility to mitigate the problems of head-of-line blocking and bidirectional transmission present in previous versions of HTTP \cite{caviglione2015deep}. 
In addition, QUIC encrypts both the transport and the user application headers preventing intermediate  nodes from accessing transport information.
While encryption provides strong guarantees of end-to-end security and confidentiality, systematic encryption of end-to-end information limits drastically network management functions and leads to performance degradation in contexts, such as the satellite network, in which QUIC may not be optimised for \cite{mami1,mami2}.

Several authors have highlighted the drawbacks of encrypting transport headers in satellite networking~\cite{quicsat,mami4}. Access to transport information allows header compression/suppression when capacity is scarce, allows to adapt the protocol behaviour to the characteristics of the satellite link (eg. HTTP acceleration, split-connections, etc.), and to implement multi-class per-hop behaviour in absence of other IP signalling. To avert losing these benefits when QUIC or other HTTP tunnels will be prevalent, a solution is urgently needed.

This paper proposes an architecture for supporting Quality of Service (QoS) in hybrid terrestrial and non-terrestrial networks \cite{bacco2019networking, cassara2022orbital} based on automated traffic classification. Traffic profiles are constructed by machine-learning (ML) algorithms using the series of packet sizes and arrival times in QUIC connections.  

A QoS architecture for satellite networks \cite{cassara2020statistical, sat.850} based on ML was first proposed in~\cite{pacheco}. The paper featured methods of feature selection, training of classifiers, and integration with the satellite resource management. Their results show that high accuracy can be achieved with a wide variety of classifiers and for a large range of traffic classes. However, feature extraction and training was still performed using transport header information (eg. the TCP port) and training samples were taken from traces over the satellite link.

When using QUIC, no transport information is available to identify a connection (even the initial connection ID  might be rotated as a result of a negotiation within the encrypted HTTP tunnel). This makes hard for Internet routers to track through deep packet inspection~\cite{dcgan}. 

Our paper shows that a high degree of accuracy can be achieved also when the transport header's information is not accessible from the classifier and the training is not done in satellite conditions (eg. with large delay or delay variance). This is good news because opens the possibility of training ML models on terrestrial networks or in laboratory conditions and then exporting the models in SATCOM equipment. Also, our results anticipate the use of satellite-connected nodes in collaborative learning environments or \textit{federated} learning.

This paper also suggests a practical implementation of the QoS architecture using a Provisioning Domain (PvD)~\cite{pvd}, which is a recent IETF standard for distribution of network configuration in trusted domains. A PvD-enabled router can include PvD containers in  IPv6 router advertisements (RAs). In this paper we argue that PvD containers can be used to include QoS information to associate IPv6 address prefixes to certain categories of traffic. Using PvD is advantageous because does not require explicit setup of a path (ie. it provides soft QoS). Rather it employs agents within the network to verify that flows conform to a given traffic profile.


The rest of the paper is organized as follows. In Section \ref{sec:relatedworks} the related works are presented. Section \ref{sec:model} discusses the general motivation of machine learning approach in the satellite context.  Section \ref{sec:scenario} provide details on the set up of the test-bed. The performance evaluation is shown in Section \ref{sec:perf}. Conclusions and recommendations are reported in Section \ref{sec:conclusions}.

%% file: sections/RelatedWorks.tex
\section{Related Work}
\label{sec:relatedworks}

The literature on Internet traffic classification using machine learning is vast (see~\cite{survey} for a recent survey on ML analysis of traffic). The proliferation of encrypted traffic in recent years resulted in an increasing use of \textit{flow-based methods} that rely on the analysis of statistical or time series features using ML. These include Naive Bayes~(NB), Support Vector Machine~(SVM), Random Forest~(RF), and K-Nearest Neighbours~(KNN)~\cite{pacheco,issues,anovel,survey2}. More recently, approaches based on deep learning and deep neural networks have appeared to classify encrypted traffic~\cite{deeppacket,datanet}.

Most of the methods in literature, however, collect features from TCP and TLS headers (eg. the TCP port), while in QUIC transport header fields are encrypted or obfuscated. A few recent papers, however, attempt to classify directly QUIC traffic.

Reference \cite{howtoachieve} echoes previous authors in saying that the most burdensome task in building an ML model for traffic classification is data labelling, which requires human intervention, whereas capture of large traces is readily available. Their approach is to use a semi-supervised method, where only a subset of flows need to be labelled to enable accurate predictions. This approach was fine tuned in~\cite{dcgan} where a deep convolutional generative adversarial network (DGCAN) was used in to classify QUIC connections. Their approach provides an accuracy of 89\% when only 10\% of the dataset was labelled.

In~\cite{cnn} authors investigate a convolutional neural network (CNN) classifier to analyse LAN traffic. The classifier is able to detect several QUIC services including Google Hangout\texttrademark, File transfer, YouTube\texttrademark and Google Play Music\texttrademark with good accuracy. While the initial analysis considers 1400 features, it is shown that through a method of features reduction, the significant set can be reduced to only three features.

In \cite{finger} traffic analysis is used to show that QUIC is prone to website fingerprinting attacks.  To build the model, an adversary can use both the vector of the number of packets for each direction and flow level statistical features including the number, size and inter-arrival time of packets. Five ML models (Random Forest, Extra Trees, K-Nearest Neighbours, Naive Bayes and SVM) are compared.

Some authors~\cite{mlqos} have successfully used ML to generate accurate user-perceived QoE (Quality of Experience) metrics for video services transported by QUIC. The accuracy of these results is an indication that ML is an adequate tool to categorise QUIC traffic.

%% file: sections/SystemArchitecture.tex
\section{A Soft QoS Architecture}
\label{sec:scenario}

In order to exploit the advantages of ML-based classification, we propose an architecture for soft coordination of end-hosts and network nodes based on Provisioning Domains (PvDs)~\cite{pvd}. End-hosts are informed about the availability of network access through Router Advertisements (RA) propagated by the local IPv6 router. RAs contain a list of network addresses that can be associated with network paths beyond the local gateway. An end-host uses one of the network addresses to forward traffic to the destination. Since the advertised addresses can be associated to descriptors of path characteristics, the end-host is made aware of the network configuration and can decide to forward traffic that matches the services offered by the network using a set of transport parameters suited for that path.

As the network nodes receive traffic from a certain address, they enforce the required treatment, but, at same time, verify that the traffic profile used by the source is consistent with the QoS definition of the path. Passive traffic analysis and identification is used to verify that the traffic generated by end-hosts effectively matches the expected set of characteristics. For instance, if a path is expected to carry progressive video streaming with a given average bitrate (e.g. DASH), the gateway should verify that the pattern of traffic resemble video streaming and the bitrate matches the expect one for that network.

A similar concept of soft QoS is used in DiffServ. In a DiffServ domain, edge routers are informed that the network supports a number of traffic classes. Edge routers initialise the DSCP (DiffServ Code Point) in the IP header before forwarding packets in the domain. The DSCP initialisation is based on a set of rules based on the IP and transport headers. Once that the packets are DSCP labelled, the internal routers use the DSCP information to enforce QoS treatments (Per Hop Behaviours). If traffic in a class exceeds the traffic envelope associated to that class, packets can be dropped or remarked (usually at the domain boundary in the \textit{policy enforcement} function). In any case, no explicit negotiation is foreseen between end-hosts and network nodes. 

The first difference between DiffServ and the present proposal is that DiffServ uses traffic classes identified by DSCPs while our architecture considers “paths” identified by IP addresses (and more precisely network identifiers) to which QoS descriptors are associated. The second difference is that, whereas there is a finite set of standard DSCPs, our architecture can associate a range of properties with paths on a much broader set using IPv6.
Associating IP prefixes with QoS classes is particularly appealing when the lower layers support virtual circuits or networks. In this case, we can identify a mapping between a \textit{L2 network} and an \textit{L3 traffic class}. The great flexibility of IPv6 can be exploited to define networks with equivalent QoS characteristics. 

This proposal employs more accurate techniques than classical token-bucket filters to characterise the traffic. For example, progressive video streaming over satellite is difficult to characterise by means of a “sustainable bit rate” due to the high burstiness. Albeit a rate-limited application, progressive video-streaming cannot be easily compared to broadcasting with constant committed information rate, which causes problems with bandwidth allocations. In addition, as the encrypted transport connections, such as QUIC, become more common, the efficacy of deep-packet inspection (DPI) is diminished as intermediaries miss critical information (such as well-known port numbers) to determine traffic classes.

%% file: sections/ProblemDefinition.tex
\section{Methodology}
\label{sec:model}

\begin{figure}[t]
\centering
\includegraphics[width=0.9\columnwidth]{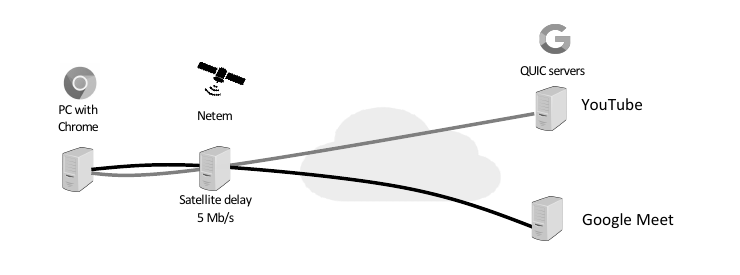}
\caption{Data collection testbed.}
\end{figure}

To determine the resilience of ML models to operating in a satellite environment, we considered a testbed for collection of two kinds of QUIC encapsulated streaming: YouTube\texttrademark~using progressive streaming over HTTP and Google Meet\texttrademark~real-time video-conferencing using MPEG-4.  The two classes of traffic were chosen not to be easily classified using deterministic features, such as the average packet size or the connection length, and therefore allow us to evaluate classifiers' ability to detect patterns in traffic.

The flows were streamed from Google servers to a local computer equipped with Wireshark~3.5, able to decrypt QUIC traces. The two streams had comparable average bitrates (about 1.2~Mb/s). A network emulator (Netem) was placed between the server and the client to change the delay, the link capacity and the delay variation. This was used to simulate   three network scenarios: A geostationary satellite with around 250~ms propagation delay in both directions, a Low-Earth-Orbit (LEO) satellite with 50~ms constant delay and 50~ms random uniform delay, and a terrestrial network without delay.

The datasets for classification were prepared by extracting from the pcap-ng traces IP throughput samples every 100~ms. Contiguous sequences of 50 or 100 samples were used as  sample patterns for the ML classification. This allowed to evaluate the performance with observation periods of 5 or 10 seconds. All the ML implementations were taken from the Python library Scikit-Learn vers.~0.24~\cite{sklearn}

Before training the ML models, the inputs were normalised dividing each sample by the difference between the minimum and maximum of the samples sequences (label "minmax" in graphs) or by using a standard normalisation ("stdnorm"). The \textit{accuracy} score was chosen as metric for comparison since class datasets (video-conferencing and progressive streaming) were roughly balanced. In a binary classification, considering a class as the positive and the other as the negative, the accuracy can be written as:
\begin{equation}
\small
    accuracy = \frac{TP+TN}{TN+TP+FP+FN}
\end{equation}
where TP and FP are the number of correct and incorrect classifications of the positive class, and TN and FN are the correct and incorrect classifications of the negative class.

Performance evaluation was carried using the classical Monte Carlo cross-validation method~\cite{montecarlo}. A single step in this method consists in randomly splitting the dataset into a \textit{training} and a \textit{validation} set (in our paper we used a proportion of 5 to 1). Then, the training set is used to fit the ML model and the validation set to calculate a sample of accuracy score. This step is repeated multiple times (200 in our setup) to estimate the distribution of accuracy scores.

\begin{figure}[t]
\includegraphics[width=\columnwidth]{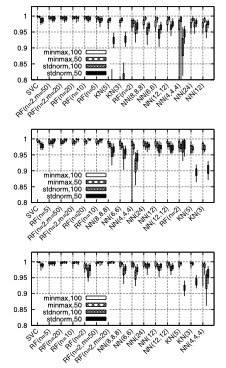}
\caption{Accuracy of ML models with different parameters for fine-tuning without delay, with delay, and delay jitter}
\label{fig:inittest}
\end{figure}

\subsection{Preliminary Tests}

An initial analysis was carried out to confirm that ML models are valid tools for traffic classification. Fig.~\ref{fig:inittest} shows the box-plot of the accuracy score for different ML models with associated \textit{hyper}-parameters in the three scenarios (respectively without delay, delay, and delay jitter). In particular, we considered the Support Vector Classifier (SVC), the Random Forest classifier (RF), multi-layer Neural Network (NN), and K-Nearest Neighbours (KNN) (definitions can be found in~\cite{survey}).

The selected hyper-parameters control the computational complexity of the algorithms. More specifically, we varied the number of layers and nodes for each layer in NN (each of the arguments in $NN(\cdot)$ is the size of a hidden layer), the depth of the trees (param.~$n$) and the number of leaves (param.~$m$) in RF, and the number of neighbours in KNN.

Fig.~\ref{fig:inittest} shows that opportunely tuning the hyper-parameters, the accuracy scores can be made very high ($>$97\%) in all scenarios. This is not surprising as reflects previous analysis~\cite{pacheco} that shows good performance as long as the model is not under-fitting. On the other hand, as complexity is increased, cross-validation usually exhibits incrementally higher performance~\cite{survey}, but these might be due to the model~\textit{over-fitting} rather than \textit{learning} the data. The next section shows how the results can generalise when the models are trained and tested in different conditions.

%% file: sections/PerformanceEvaluation.tex
\begin{figure}[t]
\includegraphics[width=\columnwidth]{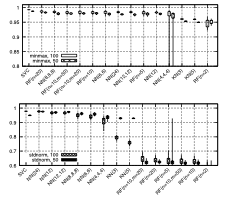}
\caption{Accuracy of ML models with different parameters: training without delay and testing with delay (250ms)}
\label{fig:crossdel}
\end{figure}

\begin{figure}[t]
\includegraphics[width=\columnwidth]{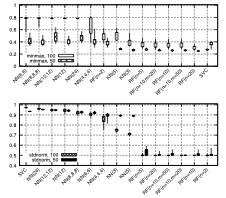}
\caption{Accuracy of ML models training with delay at 5~Mb/s and testing with bandwidth reduction (2~Mb/s)}
\label{fig:crosscap}
\end{figure}

\section{Performance Generalisation}
\label{sec:perf}

In order to generalise the results we consider in Fig.~\ref{fig:crossdel} the case where the ML algorithms were trained without satellite delay but tested with delay. Both graphs refer to a link capacity of 5~Mb/s. The top graph considers the case where the dataset is normalised using minmax, while the bottom refers to a standard normalisation. 

Since we have used a homogeneous set of features, we expected the type of normalisation to play a minor role. Instead, while the performance is only slightly reduced with respect to Fig.~\ref{fig:inittest} with minmax, the accuracy drops drastically with Random Forest when we preprocess the data using stdnorm. This suggests that tree-based methods could \textit{over-fit} the data when using throughput samples as inputs.

A similar result is also obtained when the algorithm are trained without delay and tested in the LEO scenario. In addition to the one displayed, we tested several other ML configurations, observing only marginal improvement when the complexity is further increased, but poor performance of RF. It is also worth noting that extending the observation period from 5 to 10 seconds yields limited benefits.

An even starker picture appears when the satellite capacity undergoes a significant reduction. In Fig.~\ref{fig:crosscap}, we train the ML algorithms with satellite delay and a link capacity of 5~Mb/s but we test them at 2~Mb/s. While the reduction was not such to trigger changes in transmission quality of YouTube flows, packets timings of the congestion responsive flows were significantly changed as  each of the chunks took longer to clear. As a consequence, the shape of the
throughput profile is considerably distorted in progressive streaming but a only milder distortion is present in videoconf.

The top graph referring to minmax, the drop in performance is significant across all tested configurations. For stdnorm instead, neural networks and SVC models are still able to detect the two classes of traffic, while RF and KNN performance collapse.

%% file: sections/FeaturesAnalysis.tex
\section{Feature Selection Discussion} 
\label{sec:features_amalysis}

The previous analysis considered the performance of the ML algorithms as well as their generation with a generic set of features. This section analyses a potentially more descriptive set of features and their evaluation. In particular, we used the MATLAB library MIToolbox to calculate the \textit{mutual information} and \textit{conditional entropy} to estimate the minimum set of features that prevent the model from over-fitting.

The mutual information $I(X,Y)$~\cite{cover91,eliece77} between two random variables $X$ and $Y$ measures the amount of information that the two variable shares. This means that $I(X,Y)$ measures how much knowledge from one variable can be used to reduce the error in predicting the other variable. Conversely, the conditional entropy $H(X|Y)$ \cite{cover91,eliece77} measures the amount of information needed to describe the value of the random variable $X$ knowing the value of the random variable $Y$. The mutual information is a symmetric function, the conditional entropy is not.

The set of features extracted from the time series of data traffic is shown in Table \ref{table:feature_set} for our classification model. These features, as well as the classifiers, have been proposed as the most relevant in~\cite{pacheco,survey,Tong18,zhan2021website,Zhang13}. The features in the table signed with $"*"$ belong to the essential set of features, that is, the subset that provides an accuracy for the classifiers greater than $95\%$.


\begin{table}[bt]
\centering
\caption{Extracted Features Set}
\begin{tabular}{|m{0.25cm}|>{\centering} m{1cm}|>{\centering} m{4cm}|>{\centering} m{0.75cm}|c|} 
\hline
\multicolumn{2}{|c|}{\textbf{Feature}} & \textbf{Description} & \textbf{Weight} & \textbf{Rank}\\\hline
$1)$ & $\overline{N}_{UDP}$ & Average number UDP packets sent within the window of flow & 0.0534 & $5)$\\\hline
$2)$ & $T_{W}$ & Interval time in sec. of the window of flow & 0.0044 & $10)$\\\hline
$3)$ & $Ln_{25-th}^*$ & Percentile 25-th of the UDP packet length distribution & 0.088 & $4)$\\\hline
$4)$ & $Ln_{50-th}^*$ & Percentile 50-th of the UDP packet length distribution & 0.2392 & $2)$\\\hline
$5)$ & $Ln_{75-th}$ & Percentile 75-th of the UDP packet length distribution & 0.0027 & $11)$\\\hline
$6)$ & $Ln_{90-th}$ & Percentile 90-th of the UDP packet length distribution & 0.0016 & $12)$\\\hline
$7)$ & $\Delta T_{25-th}$ & Percentile 25-th in sec. of the UDP interarrival time distribution & 0.0324 & $6)$\\\hline
$8)$ & $\Delta T_{50-th}$ & Percentile 50-th in sec. of the UDP interarrival time distribution & 0.0196 & $7)$\\\hline
$9)$ & $\Delta T_{75-th}$ & Percentile 75-th in sec. of the UDP interarrival time distribution & 0.0072 & $9)$\\\hline
$10)$ & $\Delta T_{90-th}$ & Percentile 90-th in sec. of the UDP interarrival time distribution & 0.0119 & $8)$\\\hline
$11)$ & $N_{C2S}^*$ & Number of UDP packets sent from the client toward the server & 0.3944 & $1)$\\\hline
$12)$ & $N_{S2C}^*$ & Number of UDP packets sent from the server toward the client & 0.1452 & $3$)\\\hline
\end{tabular}

\label{table:feature_set}
\end{table}


To evaluate how much through the selected features set we can predict the behaviour of the class label, we evaluate the mutual information $I(Cl,Fs)$ between the class label and the features set, and as a counter check the conditional entropy $H(Cl|Fs)$ between the class label and the feature set as inspired in \cite{cassara2021federated, MELECON}. In order to identify which of the features are essential for estimating class label values, we performed a feature analysis using the Minimum Redundancy Maximum Relevance (mRMR) method. The mRMR method selects features that are mutually distant from each other while still exhibiting a high correlation to the class label. The metrics adopted to measure the correlation between selected features and class labels can be metrics such as mutual information as in our case or functions that measure the statistical dependency. The output of the mRMR method is the subset of $i$ features, where $i$ is an input of the method, sorted following a rank of weights depending on the adopted metrics to measure the correlation between the class label and features. The weights are normalized so that the sum is one.

Given the small number of features, we crosschecked the results achieved by the mRMR method with the results obtained through an analysis of the variation of both the mutual information $I(Cl,Fs)$ and the conditional entropy $H(Cl|Fs)$ on all possible subsets of features calculated over the set of features shown in Table \ref{table:feature_set}. The features in the table generate a total number of just  $207$ subsets containing 12 features.

The last two columns in Table~\ref{table:feature_set} show the normalized weight and the  ranking, respectively, evaluated  through the analysis described above for the features in the table. The results are analyzed assuming the features clustered in the following sets time-based $\{T_W;\Delta T_{25-th};\Delta T_{50-th};\Delta T_{75-th};\Delta T_{90-th}\}$, packet-based $\{\overline{N}_{UDP};Ln_{25-th};Ln_{50-th};Ln_{75-th};Ln_{90-th}\}$, and flow-based $\{N_{C2S};N_{S2C}\}$. We face the numerical analysis aimed at showing how the clustered features belonging to the essential set affect the classifiers' performance. Precisely, we computed the possible subsets with $i$ essential features, and we removed them from the initial set. Hence, we calculate the percentage of the loss of accuracy for each classifier due to the deletion of the features. The analysis results are shown in Table~\ref{table:accuray_essential_features}. Numerical results show that filtering flow-based features impact much more than other features on degrading the classifier performance. This means that flow-based features have higher correlation with the class label, instead packet-based are weakly correlated and finally the time-based appear to be poorly correlated. The high correlation between the class label and the flow-based features is due to the more unique values these features have for the given class label than the other features. Precisely, classifier performance worsens when both the flow-based features are deleted, instead when just one of them is deleted the packet-based features are able to compensate this lack of information.

The results in Table~\ref{table:accuray_essential_features} also show that the classifiers based on SVM, RF and KNN can cope with features elimination better than the NN classifier. On the other hand, the latter performs better than the other classifiers when all flow-based features are removed. So we can assert that the classifier NN better exploits the correlation that the other essential features have with the class label. For all the tested classifiers, the tuning of the parameters has been achieved through optimized automatic procedures to guarantee a fair comparison.

\begin{table*}[t]
\centering
\caption{Percentage Decrease in Accuracy vs. Filtered Essential Features}
\begin{tabular}{|m{4.5cm}|>{\centering} m{1.3cm}|>{\centering} m{1.2cm}|>{\centering} m{1.4cm}|m{1.2cm}|} 
\hline
\centering \textbf{Deleted Features}& $\mathbf{\Delta SVC} (\%)$ & $\mathbf{\Delta RF} (\%)$& $\mathbf{\Delta KNN} (\%)$ & $\mathbf{\Delta NN} (\%)$\\\hline
$\{N_{C2S}\}$ & 0.03 & 0 & 0 & 15\\\hline
$\{Ln_{50-th}\}$ & 0 & 0 & 0 & 0.5\\\hline
$\{N_{S2C}\}$ & 0.03 & 0 & 0 & 15\\\hline
$\{Ln_{25-th}\}$ & 0 & 0 & 0 & 0.5\\\hline

$\{N_{C2S},Ln_{50-th}\}$ &0.04 & 0 & 0 & 15\\\hline
$\{N_{C2S},N_{S2C}\}$ & 71 & 75 & 72 & 15\\\hline
$\{N_{C2S},Ln_{25-th}\}$ & 0.03 & 0 & 0 & 15\\\hline
$\{Ln_{50-th},N_{S2C}\}$ &0.03 & 0 & 0 & 15\\\hline
$\{Ln_{50-th},Ln_{25-th}\}$ & 0 & 0 & 0 & 0\\\hline
$\{N_{S2C},Ln_{25-th}\}$ & 0.04 & 0 & 0 & 15\\\hline

$\{N_{C2S},Ln_{50-th},N_{S2C}\}$ & 73 & 70 & 72 & 15\\\hline
$\{N_{C2S},Ln_{50-th},Ln_{25-th}\}$ & 0.04 & 0 & 0 & 15\\\hline
$\{N_{C2S},N_{S2C},Ln_{25-th}\}$ & 66 & 69 & 83 & 15\\\hline
$\{Ln_{50-th},N_{S2C},Ln_{25-th}\}$ & 0.04 & 0 & 0 & 15\\\hline

$\{N_{C2S},Ln_{50-th},N_{S2C},Ln_{25-th}\}$ & 84 & 87 & 84 & 60\\\hline
\end{tabular}

\label{table:accuray_essential_features}
\end{table*}

Some final considerations must be made about the reasons leading to the accuracy degradation of the classifiers. From the simulations carried out, we have noticed that for the classifiers based on RF, KNN, and NN, when we filter the essential features, increase both the $FPs$ and the $FNs$, for the SVC classifier, instead increase only the $FPs$. We obtain better performance for the RF and KNN classifiers because the sum of their $FPs$ and $FNs$ is lesser than the number of $FP$ or the sum $FPs$, $FNs$  for the SVC and  NN classifier, respectively. Nevertheless, in the case of the NN classifier, this sum is almost constant during the deleting of the features. Hence, the NN classifier is more resilient to the lack of information than to the other classifiers, which confirms what discussed previously.


%% file: bare_conf.bbl
\begin{thebibliography}{10}
\providecommand{\url}[1]{#1}
\csname url@samestyle\endcsname
\providecommand{\newblock}{\relax}
\providecommand{\bibinfo}[2]{#2}
\providecommand{\BIBentrySTDinterwordspacing}{\spaceskip=0pt\relax}
\providecommand{\BIBentryALTinterwordstretchfactor}{4}
\providecommand{\BIBentryALTinterwordspacing}{\spaceskip=\fontdimen2\font plus
\BIBentryALTinterwordstretchfactor\fontdimen3\font minus
  \fontdimen4\font\relax}
\providecommand{\BIBforeignlanguage}[2]{{%
\expandafter\ifx\csname l@#1\endcsname\relax
\typeout{** WARNING: IEEEtran.bst: No hyphenation pattern has been}%
\typeout{** loaded for the language `#1'. Using the pattern for}%
\typeout{** the default language instead.}%
\else
\language=\csname l@#1\endcsname
\fi
#2}}
\providecommand{\BIBdecl}{\relax}
\BIBdecl

\bibitem{rfc9000}
J.~Iyengar and M.~Thomson, ``{QUIC: A UDP-Based Multiplexed and Secure
  Transport},'' RFC 9000, May 2021.

\bibitem{caviglione2015deep}
L.~Caviglione, N.~Celandroni, M.~Collina, H.~Cruickshank, G.~Fairhurst,
  E.~Ferro, A.~Gotta, M.~Luglio, C.~Roseti, A.~Abdel~Salam \emph{et~al.}, ``A
  deep analysis on future web technologies and protocols over broadband geo
  satellite networks,'' \emph{International Journal of Satellite Communications
  and Networking}, vol.~33, no.~5, pp. 451--472, 2015.

\bibitem{mami1}
M.~K{\"{u}}hlewind and al., ``{Challenges in Network Management of Encrypted
  Traffic},'' \emph{CoRR}, vol. abs/1810.09272, 2018.

\bibitem{mami2}
Y.~Cui, T.~Li, C.~Liu, X.~Wang, and M.~Kühlewind, ``{Innovating Transport with
  QUIC: Design Approaches and Research Challenges},'' \emph{IEEE Internet
  Computing}, vol.~21, no.~2, pp. 72--76, 2017.

\bibitem{quicsat}
N.~K. et~al., ``{QUIC: Opportunities and threats in SATCOM},'' in \emph{Proc.
  of ASMS/SPSC}, 2020, pp. 1--7.

\bibitem{mami4}
P.~A. Aranda{-}Guti{\'{e}}rrez, D.~R. L{\'{o}}pez, and T.~Fossati, ``{Analysis
  and Consideration on Management of Encrypted Traffic},'' \emph{CoRR}, vol.
  abs/1812.04834, 2018.

\bibitem{bacco2019networking}
M.~Bacco, F.~Davoli, G.~Giambene, A.~Gotta, M.~Luglio, M.~Marchese, F.~Patrone,
  and C.~Roseti, ``Networking challenges for non-terrestrial networks
  exploitation in 5g,'' in \emph{IEEE 2nd 5G World Forum (5GWF)}.\hskip 1em
  plus 0.5em minus 0.4em\relax 10.1109/5GWF.2019.8911669, 2019, pp. 623--628.

\bibitem{cassara2022orbital}
P.~Cassar{\'a}, A.~Gotta, M.~Marchese, and F.~Patrone, ``Orbital edge
  offloading on mega-leo satellite constellations for equal access to
  computing,'' \emph{IEEE Communications Magazine}, vol.~60, no.~4, pp. 32--36,
  2022.

\bibitem{cassara2020statistical}
P.~Cassar{\'a}, A.~Gotta, and T.~de~Cola, ``A statistical framework for
  performance analysis of diversity framed slotted aloha with interference
  cancellation,'' \emph{IEEE Transactions on Aerospace and Electronic Systems},
  vol.~56, no.~6, pp. 4327--4337, 2020.

\bibitem{sat.850}
\BIBentryALTinterwordspacing
N.~Celandroni, F.~Davoli, E.~Ferro, and A.~Gotta, ``Networking with
  multi-service geo satellites: cross-layer approaches for bandwidth
  allocation,'' \emph{International Journal of Satellite Communications and
  Networking}, vol.~24, no.~5, pp. 387--403, 2006. [Online]. Available:
  \url{https://onlinelibrary.wiley.com/doi/abs/10.1002/sat.850}
\BIBentrySTDinterwordspacing

\bibitem{pacheco}
F.~Pacheco, E.~Exposito, and M.~Gineste, ``A framework to classify
  heterogeneous internet traffic with machine learning and deep learning
  techniques for satellite communications,'' \emph{Computer Networks}, vol.
  173, pp. 107--213, 2020.

\bibitem{dcgan}
A.~S. Iliyasu and H.~Deng, ``Semi-supervised encrypted traffic classification
  with deep convolutional generative adversarial networks,'' \emph{IEEE
  Access}, vol.~8, pp. 118--126, 2020.

\bibitem{pvd}
D.~Anipko, ``{Multiple Provisioning Domain Architecture},'' RFC 7556, Jun.
  2015.

\bibitem{survey}
F.~Pacheco, E.~Exposito, M.~Gineste, C.~Baudoin, and J.~Aguilar, ``Towards the
  deployment of machine learning solutions in network traffic classification: A
  systematic survey,'' \emph{IEEE Communications Surveys Tutorials}, vol.~21,
  no.~2, pp. 1988--2014, 2019.

\bibitem{issues}
A.~Dainotti, A.~Pescape, and K.~C. Claffy, ``Issues and future directions in
  traffic classification,'' \emph{IEEE Network}, vol.~26, no.~1, pp. 35--40,
  2012.

\bibitem{anovel}
G.-L. Sun, Y.~Xue, Y.~Dong, D.~Wang, and C.~Li, ``{An Novel Hybrid Method for
  Effectively Classifying Encrypted Traffic},'' in \emph{In proc. of GLOBECOM
  2010}, 2010, pp. 1--5.

\bibitem{survey2}
P.~Velan, M.~\v{C}erm\'{a}k, P.~\v{C}eleda, and M.~Dra\v{s}ar, ``A survey of
  methods for encrypted traffic classification and analysis,'' \emph{Netw.},
  vol.~25, no.~5, p. 355–374, Sep. 2015.

\bibitem{deeppacket}
M.~Lotfollahi, M.~Jafari~Siavoshani, R.~Shirali Hossein~Zade, and et~al.,
  ``{Deep packet: a novel approach for encrypted traffic classification using
  deep learning},'' \emph{Soft Comput.}, no.~24.

\bibitem{datanet}
P.~Wang, F.~Ye, X.~Chen, and Y.~Qian, ``Datanet: Deep learning based encrypted
  network traffic classification in sdn home gateway,'' \emph{IEEE Access},
  vol.~6, pp. 55\,380--55\,391, 2018.

\bibitem{howtoachieve}
S.~Rezaei and X.~Liu, ``{How to Achieve High Classification Accuracy with Just
  a Few Labels: A Semi-supervised Approach Using Sampled Packets},'' in
  \emph{in proc. of ICDM 2019}, 2019.

\bibitem{cnn}
V.~Tong, H.~A. Tran, S.~Souihi, and A.~Mellouk, ``A novel quic traffic
  classifier based on convolutional neural networks,'' in \emph{2018 IEEE
  Global Communications Conference (GLOBECOM)}, 2018, pp. 1--6.

\bibitem{finger}
J.-P. Smith, P.~Mittal, and A.~Perrig, ``{Website Fingerprinting in the Age of
  QUIC},'' \emph{Proc. on PETS}, vol.~2, pp. 48--69, 2021.

\bibitem{mlqos}
M.~H. Mazhar and Z.~Shafiq, ``Real-time video quality of experience monitoring
  for https and quic,'' in \emph{IEEE INFOCOM 2018 - IEEE Conference on
  Computer Communications}, 2018, pp. 1331--1339.

\bibitem{sklearn}
F.~Pedregosa and al., ``Scikit-learn: Machine learning in {P}ython,''
  \emph{Journal of Machine Learning Research}, vol.~12, pp. 2825--2830, 2011.

\bibitem{montecarlo}
R.~R. Picard and R.~D. Cook, ``Cross-validation of regression models,''
  \emph{Jrnl. of the Am. Stat. Assoc.}, vol.~79, no. 387, pp. 575--583, 1984.

\bibitem{cover91}
J.~Cover, T.M.~Thomas, \emph{Elements of Information Theory}.\hskip 1em plus
  0.5em minus 0.4em\relax New York, NY, USA: John Wiley and Sons, Inc, 1991.

\bibitem{eliece77}
R.~McEliece, \emph{{In The Theory of Information and Coding: A Mathematical
  Framework for Communication}}, ser. Encyclopedia of Mathematics and Its
  Applications.\hskip 1em plus 0.5em minus 0.4em\relax MA, USA: Addison-Wesley
  Publishing Company: Reading, 1977, vol. Vol. 3.

\bibitem{Tong18}
V.~Tong, H.~A. Tran, S.~Souihi, and A.~Mellouk, ``A novel quic traffic
  classifier based on convolutional neural networks,'' in \emph{2018 IEEE
  Global Communications Conference (GLOBECOM)}, 2018, pp. 1--6.

\bibitem{zhan2021website}
P.~Zhan, L.~Wang, and Y.~Tang, ``Website fingerprinting on early quic
  traffic,'' 2021.

\bibitem{Zhang13}
J.~Zhang, C.~Chen, Y.~Xiang, W.~Zhou, and Y.~Xiang, ``{Internet Traffic
  Classification by Aggregating Correlated Naive Bayes Predictions},''
  \emph{Trans. on Inform. Forensics and Security}, vol.~8, no.~1, pp. 5--15,
  2013.

\bibitem{cassara2021federated}
P.~Cassar{\`a}, A.~Gotta, and L.~Valerio, ``Federated feature selection for
  cyber-physical systems of systems,'' \emph{arXiv preprint arXiv:2109.11323},
  2021.

\end{thebibliography}
